%% file: main.tex
\pdfoutput=1

\documentclass[
  fontsize=12pt,
  fleqn,
  final
]{scrartcl}

\RequirePackage{metadata/manuscript}

\begin{document}

\subfile{metadata/titlepage.tex}
\null{}

\subfile{content/sections/introduction.tex}

\subfile{content/sections/frames.tex}

\subfile{content/sections/gravities.tex}

\subfile{content/sections/inequivalence.tex}

\subfile{content/sections/bundles.tex}

\subfile{content/sections/conclusion.tex}

\subfile{content/sections/acknowledgments.tex}

\printbibliography{}

\end{document}

%% file: metadata/titlepage.tex
\thispagestyle{empty}
{
  \noindent
  \Large
  \bfseries
  About the (in){}equivalence between holonomic \textit{versus} non-holonomic theories of gravity
  \bigskip
  \bigskip
}

{
  \noindent
  \bfseries
  Guilherme Sadovski
  \bigskip
}

{
  \noindent
  \footnotesize
  Okinawa Institute of Science and Technology, 1919--1 Tancha, Onna-son, Kunigami-gun, Okinawa-ken, Japan 904--0495.
  \bigskip
}

{
  \noindent
  \footnotesize
  \rmfamily
  e-mails: \href{mailto:guilherme.sadovski@oist.jp}{guilherme.sadovski@oist.jp}.
  \bigskip
}

{
  \noindent
  \bfseries
  Abstract.\normalfont{} We investigate the scenarios in which a holonomic \textit{versus} a non-holonomic frame description of gravity theories are equivalent. It turns out that classically, the equivalence holds in a way that is independent of the particular dynamics and/or spacetime dimension. This includes general metric-affine dynamics. A global bundle-theoretical investigation is carried out, uncovering the equivalence principle as the culprit. The equivalence holds as long as the equivalence principle holds. This is not something to be expected when non-invertible configurations of the \textit{vielbein} field are taken into account. In such case, the gauge-theoretical description of gravity \textquote{unsolders} from spacetime, and one has to decide if gravity is spacetime geometry or an internal gauge theory.
  \bigskip
}

{
  \noindent
  \rule{\textwidth}{1pt}
  \vspace{-4.5ex}
  \tableofcontents
  \noindent
  \rule{\textwidth}{1pt}
}

%% file: content/sections/introduction.tex
\section{Introduction}\label{sec:introduction}

General Relativity (GR) is based on A. Einstein's minimalistic assumption that all degrees of freedom of the gravitational field can be encoded in a single tensorial field: the metric field $g$. This historically attracted immediate criticism. Most notably from E. Cartan, who strongly advocated that metrical and affine structures are two logically distinct concepts. In Cartan's point of view, we should minimize \textit{ad hoc} assumption about the spacetime manifold. Thus, the degrees of freedom of the gravitational field should be described, in generality, by $g$ and an affine connection $\Gamma$ as independent fundamental fields.

Today, more than 100 year later, many so-called alternatives to GR have been developed. Most of them falling under the umbrella of metrical (GR, $f\left(R\right)$, \textit{etc.}), scalar-metrical (Brans-Dicke, Horndeski, \textit{etc.}), vector-metrical (Will-Nordtvedt, Hellings-Nordtvedt, \textit{etc.}), bimetrical (Rosen, Rastall, \textit{etc.}), affine (Teleparallel, Symmetric Teleparallel, \textit{etc.}), metric-affine (Einstein-Cartan (EC), metric-affine $f(R)$, \textit{etc.}), or gauge-theoretical (Einstein-Cartan-Sciama-Kibble (ECSK), affine gauge theory, \textit{etc.}). Each of them claiming their own set of fundamental variables.

In teleparallelism~\cite{aldrovandi2013,maluf2013}, there exists a single fundamental field: a metric-compatible flat $\Gamma$. Gravity, in this formulation, is exclusively a manifestation of non-vanishing torsion. In symmetric teleparallelism~\cite{ferraris1982a,nester1999}, it is a torsion-free flat $\Gamma$, and gravity is exclusively manifestation of non-vanishing non-metricity. In EC theory~\cite{hehl1974,hehl1976a,trautman2006}, metric-affine $f(R)$~\cite{sotiriou2007,sotiriou2009,sotiriou2010,olmo2011}, generic metric-affine~\cite{hehl1976b}, \textit{etc.}, $g$ and $\Gamma$ are the two independent fundamental fields (Cartan's philosophy). Gravity, in these, is a manifestation of the non-vanishing curvature, torsion, and/or non-metricity of $\Gamma$.

In the gauge-theoretical branch~\cite{ivanenko1983a,sardanashvily2016a,hehl1995,gronwald1995,gronwald1997,sardanashvily2002a,hehl2012}, $g$ and $\Gamma$ are effective rather than fundamental fields. This approach mainly consists of gauge theories for external symmetries --- usually the general affine group $Aff\left(n\right)$, one of its subgroups, or their supersymmetric extensions --- and, in general, have a soldering form $e$ and a gauge connection $A$ as set of fundamental fields~\cite{hehl1995}. Historically, the Lorentz group $SO\left(1,3\right)$ was the first external group to be gauged, giving raise to the ECSK theory~\cite{utiyama1956,kibble1961,sciama1962,sciama1964}. The generators of $\mathfrak{so}\left(1,3\right)$ are antisymmetric, which makes Lorentz connections metric-compatible. In ECSK theory, gravity is exclusively a manifestation of curvature and/or torsion.

Recently, the gauging of $Spin(4)$ have also shown to produce gravitation~\cite{weldon2001,lippoldt2014,emmrich2022}. In spin base invariant models, the set of fundamental fields consists of Dirac matrices $\gamma$ (in substitution of $e$), satisfying the Clifford algebra $\left\{ \gamma, \gamma \right\} = 2g$, and an $Spin(4)$ gauge connection $\hat A$. Analogously to ECSK theory, the generators of $\mathfrak{spin}(4)$ are antisymmetric, which makes $\hat A$ metric-compatible as well. In spin base invariant models, gravity is a manifestation of non-vanishing curvature and/or torsion of $\hat A$.

As one can see, the \textquote{Einstein \textit{versus} Cartan debate} is pretty much alive~\cite{hehl1974,fay2007,capozziello2011,nojiri2011,clifton2012,berti2015,iosifidis2020,golovnev2022}. And, in despite of the very recent advances in observational physics --- with emphasis in very long baseline interferometry and multi-messenger astronomy ---, a concrete answer seems unlikely in the near future~\cite{lammerzahl1997,obukhov2014,broderick2014,yagi2016,baker2017,mizuno2018,sunny2019,eht2019,bahamonde2021,cantata2021,lobo2021,ferreira2022}. Under this light, a classification of all these distinct description is of upmost importance.

Classically, GR, Teleparallel and Symmetric Teleparallel are known to be equivalent among themselves~\cite{jimenez2019,capozziello2022a}. As we review in more details in Section~\ref{sec:gravities}, EC theory, in the presence of matter carrying vanishing hypermomentum currents, is also equivalent to GR~\cite{hehl1974}. Analogously, metric-affine $f(R)$ in vacuum is equivalent to GR~\cite{sotiriou2007}. This last result can strike as quite surprising, given that the only case in which metrical $f(R)$ is equivalent to GR is when $f(R)=R$~\cite{olmo2007}. If $\phi \equiv f'(R)$ is an invertible field transformation\footnote{The prime $'$ indicates derivative of the function with respect to its argument.}, then metrical $f(R)$ is equivalent to the scalar-metrical $\omega=0$ Brans-Dicke theory with potential $V(\phi) \equiv R(\phi)\phi-f(R(\phi))$~\cite{teyssandier1983}. Under the same field transformation, metric-affine $f(R)$, in the presence of matter carrying vanishing hypermomentum currents, is equivalent to $\omega=-3/2$ Brans-Dicke theory with potential $V(\phi)$~\cite{sotiriou2009}. These $f(R)$ (in){}equivalences smoothly hold in the limit $f(R)\rightarrow R$~\cite{olmo2007}.

Many gauge-theoretical models of gravity have classical equivalence to metrical, affine or metric-affine models. For instance, the gauge theory for spacetime translations ($\mathbb{R}^4$), first developed in~\cite{hayashi1967}, was shown to be equivalent to teleparallelism in~\cite{cho1976,hayashi1977,hayashi1979} --- see~\cite{aldrovandi2013} for a historical account. The ECSK theory, on the other hand, was born as a gauge theory of gravity and is equivalent to EC theory. Spin base invariant gravity (with vanishing spin torsion) and ECSK theory (in the presence of matter with vanishing hypermomentum currents) are both equivalent to GR\@. In fact, these equivalences are so widely spread in physics literature, that they are commonly referred to as just different formalisms of a same underlying physical theory: the metrical or holonomic $(g,\Gamma)$ \textit{versus} the \textit{vielbein} or non-holonomic $(e,A)$\footnote{Or $(\gamma,\hat A)$, in the case of spin base invariant gravity.}.

In this work, we avoid phrasing holonomic $(g,\Gamma)$ \textit{versus} non-holonomic $(e,A)$ theories of gravity as just different formalisms. We do not take the aforementioned equivalences for granted, and we adopt the point of view that holonomic \textit{versus} non-holonomic theories of gravity are fundamentally distinct --- unless unequivocally proven otherwise. The reason for this is three-fold: (i) a classical equivalence might not hold quantum mechanically; (ii) to the author's knowledge, it is not known if this equivalence hold for a general metric-affine dynamics and; (iii) it is known to fail on degenerate spacetimes~\cite{kaul2016a,kaul2016b,kaul2019}.

As one can imagine, point (i) is tricky to be addressed, as we do not know how to properly formulate a quantum theory of gravity. However, attempts have been made within a path integral approach. In~\cite{lippoldt2014}, it was shown that the field transformation given by $\left\{\gamma,\gamma\right\}=2g$, introduces a trivial factor in the functional measure $\mathcal{D}g$ (of quantum GR), if transformed from the functional measure $\mathcal{D}\gamma$ (of quantum spin base invariant gravity with vanishing spin torsion). In~\cite{zanelli2003}, the quantum equivalence between ECSK theory in vacuum and GR is also established. It, however, ignores non-globally hyperbolic spacetimes and the particular ghost structure of each theory\footnote{The former is invariant under $Diff(4) \ltimes SO(1,3)$, while the latter is under $Diff(4)$.}. Reference~\cite{dario2011} shows that such oversights are irrelevant at 1-loop level. Nonetheless, functional renormalization group (FRG) analyses show important qualitative differences in the $g$ \textit{versus} $e$ theory space, and their respective FRG flows~\cite{reuter2010}. More concretely in~\cite{reuter2012}, the ghost fields associated to the local Lorentz invariance are shown to contribute quite significantly to the running of the Newton and cosmological constant in the non-perturbative regime\footnote{This also should not be seen as a definite answer, as the FRG framework relies on a specific truncation and on the use of a Euclidean signature for the metric.}~\cite{reuter2012}. As one can see, the quantum equivalence is pretty much an open issue even in the simplest models --- further discussions can be found in~\cite{reuter2013,reuter2015,reuter2016}.

Point (ii) is addressed in Section~\ref{sec:on-shell_equivalence}, and is the main original contribution of this present work. We establish the equivalence between holonomic \textit{versus} non-holonomic gravity theories, in a manner independent of any particular metric-affine dynamics and/or spacetime dimensions. This equivalence holds as long as the field transformations in~\eqref{eq:field-transformation} hold. In Section~\ref{sec:geometric_picture}, we clarify the global bundle-theoretical nature of these field transformations and their relation to the equivalence principle. It happens that the former is connected to the latter via a non-degenerate $e$. This brings us to point (iii). It refers to regions of spacetime where $e$ or, equivalently, the \textit{vielbein} field $\tensor{e}{^A_\mu}\left(x\right)$ --- its matrix representation --- is non-invertible. As we discuss in more details in Section~\ref{sec:geometric_picture} and~\ref{sec:conclusions}, on these regions, the gauge-theoretical description of gravity unsolders from spacetime, and all the gauge-theoretical equivalences aforementioned are expected to fail.

Sections~\ref{sec:holonomic_and_non-holonomic_frames}~and~\ref{sec:gravities} serve as prelude to the generalizations addressed in Section~\ref{sec:on-shell_equivalence}. They contain a detailed review on holonomic \textit{versus} non-holonomic variables, and the reasons for the classical equivalence among GR, EC and ECSK theory in non-degenerate spacetimes. Finally, in what follows, we consider the category of smooth $n$-dimensional manifolds ($C^\infty$ $n$-manifolds), unless stated otherwise. Greek, lower-case and upper-case Latin letters range from $0$ to $n-1$, unless stated otherwise.

%% file: content/sections/frames.tex
\section{Frames}\label{sec:holonomic_and_non-holonomic_frames}

A frame is generally considered as an ordered set of linearly independent vectors spanning some vector space. Given such general definition, one can image the multitude of different frames one could potentially define over a point $x$ of a manifold $X$. Or, the multitude of fields of frames one could potentially define over a neighborhood $U\subseteq X$ of $x$. Such fields are what E.~Cartan first described in generality in~\cite{weyl1938} as \textquote{moving frames} over $U$.

\subsection{Holonomic}\label{ssec:holonomic}

The most commonly defined frame is the so-called coordinate or holonomic frame. Sometimes also called \textquote{world} or \textquote{spacetime} frame for reasons that will become clear in a moment. At $x$, it is defined as a particular choice of ordered basis for $T_x X$ --- the tangent space of $X$ at $x$. Such choice consists of derivations\footnote{Derivations at a point $x\in X$ are linear maps $D_x:C^{\infty}\left(X\right)\rightarrow \mathbb{R}$ acting on smooth functions on $X$ and satisfying $D_x \left(f\circ g\right)=D_x f g(x) + f(x)D_x g$.} of the kind
\begin{equation}
  \label{eq:holonomic-basis}
  \partial_\mu|_x \left[f\right] \equiv \frac{\partial}{\partial x^\mu} \left(f\circ\phi^{-1}\right)|_{\phi(x)}\;,
\end{equation}
acting on smooth functions $f: U \rightarrow \mathbb{R}$ on $U$. Here, $ \phi: U \rightarrow A \subseteq \mathbb{R}^n $ is a local chart giving an ordered set of $n$ Euclidean labels, $\phi(x)=\left\{x^0,\ldots,x^{n-1}\right\}$, to each point $x$ in $U$ --- $x^\mu$ represents each of such labels.

We can  write down a holonomic frame at $x$ as the ordered set $\left\{\partial_0|_x, \ldots, \partial_{n-1}|_x\right\}$. When the context is sufficiently clear, we might refer to it simply by its elements $\partial_\mu|_x$ and \textit{vice-versa}. Moreover, a field of holonomic frames over $ U $ is a map uniquely assigning to each $ x \in U $ a frame $ \partial_\mu|_x $. Such field, henceforth denoted as $\partial_\mu \left(x\right)$, is what Cartan would call a tangent (or holonomic, in our language) moving frame.

Another very commonly defined frame is the so-called holonomic co-frame. It is defined using functionals acting on $T_x X$ and spanning $T_x^{*} X$ --- the co-tangent space of $X$ at $x$. The natural choice is $dx^\mu|_x$, implicitly given by the relation $dx^\mu|_x \left(\partial_\nu|_x\right)=\tensor{\delta}{^\mu_\nu}$. Let the ordered basis $\left\{dx^0|_x,\ldots,dx^{n-1}|_x\right\}$ of $T^*_x X$ be one such co-frame at $x$, henceforth denoted by $dx^\mu|_x$. A field of holonomic co-frames over $U$ is denoted by $dx^\mu(x)$, and we call it a holonomic moving co-frame.

Holonomic frames explicitly use the concept of a local chart in their definition. As a result, they --- and, consequentially, their co-frames --- have a unique behavior. Consider another chart, $ \phi': U \rightarrow A' \subseteq \mathbb{R}^n $, giving different Euclidean labels, $ \phi'(x) = \{ x^{0'}, \ldots, x^{ \left( n-1 \right)' } \} $, to the same region $ U $ in $ X $. The composition map $ \phi' \circ \phi^{-1}: A \rightarrow A' $, known as a transition function on $U$, represents for $x$ the change of labels $ x^\mu \mapsto x^{ \mu' } $. If such change happens to be bijective and smooth $ \forall \; x \in U$, then holonomic frames --- and their co-frames --- all over this region suffer the action of an $ n \times n $ invertible matrix. Namely, for moving frames,
\begin{equation}
  \label{eq:holonomic-frame-transf-rule}
  \partial_{\nu'}(x) = \tensor{J}{^\mu_{\nu'}}\left(x\right) \partial_\mu(x) \quad ; \quad \tensor{J}{^\mu_{\nu'}}\left(x\right) \equiv \partial_{\nu'} x^\mu \;,
\end{equation}
and, for their co-frames,
\begin{equation}
  \label{eq:holonomic-coframe-transf-rule}
  dx^{\nu'}(x) = \tensor{J}{^{\nu'}_\mu}\left(x\right)dx^{\mu}(x) \quad ; \quad \tensor{J}{^{\nu'}_\mu}\left(x\right) \equiv \partial_\mu x^{\nu'} \;,
\end{equation}
where $\tensor{J}{^{\nu'}_\mu}\left(x\right)$ is the Jacobian matrix of $\phi'\circ {\phi}^{-1}|_{\phi(x)}$ and $\tensor{J}{^\mu_{\nu'}}(x)$ is its inverse. In other words, holonomic frames and their co-frames are sensible to changes of local charts in $X$. If this change occurs in an invertible and, at least $C^1$ manner, then the corresponding change in each $T_x U$ can be seen as an action of the general linear group $GL\left(n,\mathbb{R}\right)$. In Physics, this intimate relationship between holonomic frames and the base manifold --- usually spacetime --- is the reason why they are also called \textquote{world} or \textquote{spacetime} frames.

\subsection{Non-holonomic}\label{ssec:non-holonomic}

In contrast, generic frames do not rely on $\phi$ for their definition. Unlike~\eqref{eq:holonomic-basis}, most frames are not sensible to changes of charts in $X$. This majority is referred to as non-holonomic, and they are naturally present in physical theories with or without gravity. In the study of a quantized Dirac field over a fixed spacetime 4-manifold, a relevant non-holonomic moving frame is defined by a map giving to each event an ordered orthonormal basis in $\mathbb{C}^4$. Physically, each of these frames can be interpreted as a Stern-Gerlach experimental apparatus, able to measure the spin orientation of the fundamental excitations of the Dirac field --- particle and anti-particle --- at a specific point in space and time.

In the case of pure relativistic theories of gravity, which is the focus of the present work, non-holonomic moving frames are brought into light by the geometric equivalence principle~\cite{ivanenko1983a,sardanashvily1994a,sardanashvily2016a}. Consider the moving frame $\tau_a(x)$ on $ U \subseteq X $ that uniquely associates to each $ x \in U $ the ordered basis $ \tau_a|_x $ spanning the vector space $ V_x $. We want $ V_x $ to carry a linear action of $ SO \left( 1, n-1 \right) $ --- the isometry group of the $n$-dimensional Minkowski space, $ M $. In other words, $ \tau_a(x) $ over $U$ transforms according to
\begin{equation}
  \label{eq:lorentz_frames_transf}
  {\tau}_{a'}(x) = \tensor{\Lambda}{^b_{a'}}(x)\tau_b(x) \;,
\end{equation}
where $\tensor{\Lambda}{^b_{a'}}(x)$ is a matrix representation of $SO(1,n-1)$. In principle, equation~\eqref{eq:lorentz_frames_transf} is not the result of any local change of chart on $X$. Thus, it is non-holonomic in nature. It, however, can be interpreted as holonomic in $M$. If we forget its $x\in X$ dependence, we have the right to interpret~\eqref{eq:lorentz_frames_transf} as the result of a change of global charts in $M$ that preserves the globally defined Minkowski metric $\eta$ there. And there, $\tau_a$ as well as $\tau_{a'}$ are global (and constant) holonomic moving frames, orthogonal with respect to (w.r.t.) $\eta$. This is exactly the kind of moving frames in which Special Relativity (SR) is defined.

Under this point of view, the equivalence principle is fulfilled in a generic curved geometry if it is possible to define a $\tau_a(x)$ on every possible $ U \subseteq X $. This effectively covers all of $X$ with frames in which Lorentz invariants can be defined. Physically, they carry exactly the same meaning as in SR\@: a force-free clock and $n-1$ linearly independent rods at each point in spacetime\footnote{ Unfortunately, the equivalence principle is vaguely defined throughout history and literature. There are many (often inequivalent) ways to state it physically and/or mathematically~\cite{ivanenko1983a,sardanashvily1994a,dicasola2015a,paunkovic2022a,capozziello2024a}. In this work, we adopt the so-called \textquote{geometric} version because it is the one naturally related to the issue at hand --- \textit{vide} Section~\ref{ssec:the_equivalence_principle}. Additionally, this version is physically and mathematically unambiguous as it builds upon the gauge principle, the Chern-Weil theory and the Erlangen program. }.

One can go ahead and quickly define the co-frame $\tau^a|_x$ of $\tau_a|_x$ as the functionals acting on $V_x$ and spanning $V_x^*$ such that $\tau^a|_x \left(\tau_b|_x\right)=\tensor{\delta}{^a_b}$. It transforms non-holonomically according to
\begin{equation}
  \label{eq:lorentz_coframes_transf}
  {\tau}^{a'}(x) = \tensor{\Lambda}{^{a'}_b}(x)\tau^b(x) \;,
\end{equation}
where $\tensor{\Lambda}{^b_{a'}}(x)$ is the inverse matrix of $\tensor{\Lambda}{^{a'}_b}(x)$.

We finally conclude this section with two remarks: (i) the discussion above is very familiar. Equations~\eqref{eq:holonomic-frame-transf-rule} and~\eqref{eq:holonomic-coframe-transf-rule} are nothing but the transformation rules of covariant and contravariant coordinate vectors, respectively, exhaustively discussed throughout the literature. Equations~\eqref{eq:lorentz_frames_transf} and~\eqref{eq:lorentz_coframes_transf} should also look ordinary --- very similar to the transformation laws that the 1-form \textit{vielbein} $e^a(x)\equiv \tensor{e}{^a_\mu}\left(x\right)dx^\mu$ and its inverse field obey. However, the relation between $\tau^a(x)$ and $e^a(x)$ is a bit more subtle than just an equality, which leads us to the second remark; (ii) these quantities and relations belong to vector bundle structures over $X$~\cite{trautman1970}. We enter in more details about this in Section~\ref{sec:geometric_picture}. For the moment, we blindly use the fact that the non-holonomic group $SO(1,n-1)$ can be enlarged to a non-holonomic $GL(n,\mathbb{R})$ without compromising any aspect of the discussion above --- including the validity of the equivalence principle. This allows to extend the discussion in Section~\ref{sec:gravities} to very general theories of gravity --- not just GR or its torsional extensions. To emphasize this change, capital Latin letters represents non-holonomic $GL(n,\mathbb{R})$ indexes while Greek letters stay exclusively to holonomic $GL(n,\mathbb{R})$ ones.

%% file: content/sections/gravities.tex
\section{Gravity theories}\label{sec:gravities}

\subsection{Cartan's philosophy}\label{ssec:cartan's_philosophy}

In holonomic moving frames, a fundamental ingredient of the Einstein-Palatini approach is the understanding that a metric tensor field $g(x)$ and an affine connection $\nabla$ are \textit{\`a priori} two logically distinct concepts over a spacetime region $ U \subseteq X $. Physically, $g(x)$ introduces how a set $ \partial_\mu (x) $ of observers on $ U $, can perform \textquote{dot product} measurements on pairs of vector fields. Indeed, $g\left[\partial_\mu,\partial_\nu\right] \left(x\right)\equiv g_{\mu\nu} \left(x\right)$ is a smooth function on $U$ associating a set of $n\left(n+1\right)/2$ real numbers to each $ x \in U $. Such numbers are interpreted as \textquote{sizes} ($\mu=\nu$) and \textquote{angles} ($\mu\neq \nu$) between $\partial_\mu$ and $\partial_\nu$\footnote{From now on, whenever the context is sufficiently clear, the $x$ dependence of fields will be omitted.}. Meanwhile, $\nabla$ introduces a recipe on how vector fields can be differentiated along others. Indeed, $ dx^\alpha \left( \nabla \left[ \partial_\mu, \partial_\beta \right] \right) \equiv \tensor{\Gamma}{^\alpha_{\beta\mu}} $ is another smooth function on $ U $ associating a set of $n^3$ real numbers for each $ x \in U $, in general. Such numbers are interpreted as the infinitesimal variation of $ \partial_\beta $ along $\partial_\mu$ as measured by $dx^\alpha$ at each $x$.

It is true, however, that there is a canonical way to collapse affine concepts into metrical ones. For instance, to impose that the only way $\partial_\beta$ can vary along $\partial_\mu$ is by an infinitesimal \textquote{rotation}, \textit{i.e.}, a specific first-order change in $g_{\beta\mu}$. Such change is what is known as Christoffel symbols. A detailed analysis of the irreducible decomposition of $\tensor{\Gamma}{^\alpha_{\beta\mu}}$ reveals that this is equivalent to forbid $\partial_\beta$ to pick up any infinitesimal shear, dilatation and/or displacement variations along $\partial_\mu$. Or, equivalently, that $\tensor{\Gamma}{^\alpha_{\beta\mu}}$ satisfies the following constraints:
\begin{subequations}\label{eq:vanishing_torsion_nonmetricity}
  \begin{align}
    \tensor{T}{^\alpha_{\beta\mu}} & \equiv \tensor{\Gamma}{^\alpha_{\beta\mu}}-\tensor{\Gamma}{^\alpha_{\mu\beta}} = 0 \label{eq:vanishing_torsion} \;,                                                       \\
    \tensor{Q}{_{\alpha\beta\mu}}  & \equiv \partial_\alpha g_{\beta\mu} -\tensor{\Gamma}{^\nu_{\beta\alpha}}g_{\nu\mu}-\tensor{\Gamma}{^\nu_{\mu\alpha}}g_{\beta\nu} = 0 \label{eq:vanishing_nonmetricity}\;,
  \end{align}
\end{subequations}
where $\tensor{T}{^\alpha_{\beta\mu}}$ and $Q_{\alpha\beta\mu}$ are the so-called torsion and non-metricity tensor fields, respectively. As one can see, this is indeed a very particular and constrained geometry, first formulated by B. Riemann in 1854\footnote{This work was never published by Riemann himself. He first laid out the foundational aspects of what is now called a Riemannian $n$-manifold in a lecture, \textit{On the hypothesis that lie at the foundation of geometry}, at G\"{o}ttingen University, as part of the qualification process for him to become a \textit{Privatdozent} (lecturer).}, and fully embodied in GR 60 years later\@.

On the other hand, other geometries such as Weitzenb\"{o}ck ($R=0, \; T\neq 0, \; Q=0$), Weyl ($R=0, \; T=0, \; Q\neq 0$), Riemann-Cartan  ($R\neq 0, T\neq 0$, $Q=0$) or metric-affine ($R\neq 0, T\neq 0$, $Q\neq 0$) are as compatible with current observational data as the Riemannian hypothesis~\cite{hehl1974,lammerzahl1997,fay2007,capozziello2011,nojiri2011,clifton2012,obukhov2014,broderick2014,berti2015,yagi2016,baker2017,mizuno2018,sunny2019,eht2019,jimenez2019,iosifidis2020,bahamonde2021,cantata2021,lobo2021,ferreira2022,golovnev2022,capozziello2022a}.\ \textit{Em prol} of generality, and advocating Cartan's philosophy, we consider $g_{\mu\nu}$ and $\tensor{\Gamma}{^\alpha_{\beta\mu}}$ as completely independent fields, each carrying part of the classical degrees of freedom of gravity. Unless, dynamically stated otherwise via the field equations.

Following the above reasoning, the curvature tensor field
\begin{equation}
  \label{eq:curvaturetensor}
  \tensor{R}{^\alpha_{\beta\mu\nu}} = \partial_\mu\tensor{\Gamma}{^\alpha_{\beta\nu}}-\partial_\nu\tensor{\Gamma}{^\alpha_{\beta\mu}}+\tensor{\Gamma}{^\alpha_{\rho\mu}}\tensor{\Gamma}{^\rho_{\beta\nu}}-\tensor{\Gamma}{^\alpha_{\rho\nu}}\tensor{\Gamma}{^\rho_{\beta\mu}} \;
\end{equation}
should be considered as a function of $\tensor{\Gamma}{^\alpha_{\beta\mu}}$ and its derivatives alone --- not of $g_{\mu\nu}$ and its derivatives. This tensor differs from the Riemann curvature since, again, we are not making any assumption on how $\tensor{\Gamma}{^\alpha_{\beta\mu}}$ differs from Christoffel symbols.

\subsection{EC and ECSK theory}\label{ssec:ec_theory}

The dynamics of $n$-dimensional EC theory is defined by the $n$-dimensional Einstein-Hilbert-Palatini (EHP) action,
\begin{equation}\label{eq:holonomic-EP-action}
  S_{\text{EHP}}\left[g_{\mu\nu},\;\tensor{\Gamma}{^\alpha_{\beta\mu}}\right]=\int_{X} d^n x \sqrt{-g}\tensor{R}{^\alpha_{\mu\alpha\nu}}g^{\mu\nu}\;.
\end{equation}
where $g_{\mu\nu}$ corresponds to a Lorentzian metric, while $g$ is its determinant and $g^{\mu\nu}$ is its inverse. The field equations obtained from the functional variation w.r.t. $g_{\mu\nu}$ and $\tensor{\Gamma}{^\alpha_{\beta\mu}}$ are, respectively,
\begin{subequations}\label{eq:holonomic-field-eqs}
  \begin{align}
    -\sqrt{-g}G^{\mu\nu}                                                                                                                                                                                            & = 0 \;, \label{eq:holonomic-einstein-like-field-eqs} \\
    -\sqrt{-g}\left[\tensor{T}{^\mu_\alpha^\beta}-\tensor{Q}{_\alpha^{\beta\mu}} + \frac{1}{2} \left(g^{\beta\mu}\tensor{Q}{_{\alpha\nu}^\nu}+\tensor{\delta}{_\alpha^\mu}\tensor{Q}{^\beta_\nu^\nu}\right) \right] & = 0\;, \label{eq:holonomic-cartan-like-field-eqs}
  \end{align}
\end{subequations}
where $G_{\mu\nu} \equiv \tensor{R}{^\alpha_{\mu\alpha\nu}} - \frac{1}{2} \tensor{R}{^\alpha_{\beta\alpha\lambda}}g^{\beta\lambda}g_{\mu\nu}$ is the post-Riemannian Einstein tensor;  asymmetric in $\mu\nu$.

The $n$-dimensional ECSK theory, which is formulated in non-holonomic frames, has its dynamics defined by the so-called \textit{Vielbein}-Einstein-Palatini (VEP) action,
\begin{equation}
  \label{eq:vep-action}
  S_{\text{VEP}} \left[\tensor{e}{^A_\mu}, \tensor{A}{^A_{B\mu}}\right] = \int_X d^n x e\tensor{F}{^A_{B\mu\nu}}\tensor{e}{^\mu_A}\tensor{e}{^{B\nu}}  \;,
\end{equation}
where $\tensor{e}{^A_\mu}$ is the \textit{vielbein} field, $e$ is its determinant and $\tensor{e}{^\mu_A}$ is its inverse satisfying
\begin{subequations}
  \begin{align}
    \tensor{e}{^A_\mu}\tensor{e}{^\mu_B} & = \tensor{\delta}{^A_B}\;, \label{eq:inversepullbackvielbein} \\
    \tensor{e}{^A_\mu}\tensor{e}{^\nu_A} & = \tensor{\delta}{^\nu_\mu} \label{eq:inversevielbein}\;.
  \end{align}
\end{subequations}
Under a non-holonomic $GL\left(n,\mathbb{R}\right)$ action, they transform as
\begin{subequations}
  \begin{align}
    \label{eq:active_transf_vielbein}
    \tensor{e}{^{A'}_\mu} & = \tensor{\Lambda}{^{A'}_{B}} \tensor{e}{^B_\mu} \;, \\
    \tensor{e}{^\mu_{A'}} & = \tensor{\Lambda}{^{B}_{A'}} \tensor{e}{^\mu_B} \;,
  \end{align}
\end{subequations}
while under a holonomic one, they do as
\begin{subequations}
  \begin{align}
    \label{eq:passive_transf_vielbein}
    \tensor{e}{^A_{\nu'}} & = \tensor{J}{^\mu_{\nu'}} \tensor{e}{^A_\mu} \;, \nonumber \\
    \tensor{e}{^{\nu'}_A} & = \tensor{J}{^{\nu'}_{\mu}} \tensor{e}{^\mu_A} \;.
  \end{align}
\end{subequations}
Furthermore, $\tensor{A}{^A_{B\mu}}$ is a $GL\left(n,\mathbb{R}\right)$ connection and its curvature,
\begin{equation}
  \label{eq:nonholcurvature}
  \tensor{F}{^A_{B\mu\nu}}=\partial_\mu\tensor{A}{^A_{B\nu}}-\partial_\nu\tensor{A}{^A_{B\mu}}+\tensor{A}{^A_{C\mu}}\tensor{A}{^C_{B\nu}}-\tensor{A}{^A_{C\nu}}\tensor{A}{^C_{B\mu}} \;,
\end{equation}
is, unmistakably, a function of connection $\tensor{A}{^A_{B\mu}}$ and its derivatives alone.

The reader might notice that the above fields are being called non-holonomic when they clearly also have holonomic indexes. As we clarify in Section~\ref{sec:geometric_picture}, these fields are local projections on $X$ of truly non-holonomic quantities living on internal bundles. Due to this mixed nature, $g_{\mu\nu}$ and its inverse are still present in the non-holonomic framework --- albeit not as a dynamical field. Additionally, an extra metric, $g_{AB}$, and its inverse $g^{AB}$, are also present in order to \textquote{rise} and \textquote{lower} non-holonomic indexes\footnote{The relation between these two metrics (equation~\eqref{eq:metrics}) is addressed in Section~\ref{sec:geometric_picture}.}.

The field equations obtained from the functional variation w.r.t. $\tensor{e}{^A_\mu}$ and $\tensor{A}{^A_{B\mu}}$ are, respectively,
\begin{subequations}\label{eq:nonholonomic-field-eqs}
  \begin{align}
    e\delta^{\mu\nu\lambda}_{\alpha\beta\gamma}\tensor{e}{^\alpha_A}\tensor{e}{^\beta_B}\tensor{e}{^\gamma_C}\tensor{F}{^{BC}_{\nu\lambda}}                                                                                                          & = 0 \;, \label{eq:nonholonomic-einstein-like-field-eqs} \\
    e \left(\delta^{\mu\nu\lambda}_{\alpha\beta\gamma}\tensor{e}{^\alpha_A}\tensor{e}{^{B\beta}}\tensor{e}{^\gamma_C}\tensor{T}{^C_{\nu\lambda}}-2\delta^{\mu\nu}_{\alpha\beta}\tensor{e}{^\alpha_A}\tensor{e}{^\beta_C}\tensor{Q}{_\nu^{BC}}\right) & = 0 \;, \label{eq:nonholonomic-cartan-like-field-eqs}
  \end{align}
\end{subequations}
where some useful definitions were used\footnote{\setlength{\abovedisplayskip}{-6pt}
  \begin{align*}
    \delta^{\mu_1 \cdots \mu_p}_{\nu_1\cdots \nu_p} & \equiv \frac{1}{\left(n-p\right)!}\epsilon^{\mu_1 \cdots \mu_p \lambda_{p+1}\cdots\lambda_{n}}\epsilon_{\nu_1\cdots\nu_p\lambda_{p+1}\cdots\lambda_n}\;,     \\
    \tensor{T}{^A_{\mu\nu}}                         & \equiv \partial_\mu \tensor{e}{^A_\nu}- \partial_\nu \tensor{e}{^A_\mu} + \tensor{A}{^A_{B\mu}}\tensor{e}{^B_\nu}-\tensor{A}{^A_{B\nu}}\tensor{e}{^B_\mu}\;, \\
    \tensor{Q}{_\mu^{AB}}                           & \equiv \tensor{A}{^{AB}_\mu}+\tensor{A}{^{BA}_\mu} \;.
  \end{align*}}. In particular, $\delta^{\mu_1\cdots \mu_p}_{\nu_1\cdots\nu_p}$ is the generalized Kronecker delta, $\epsilon_{\mu_1\cdots \mu_n}$ is the permutation symbol, $\tensor{T}{^A_{\mu\nu}}$ is the non-holonomic torsion and $\tensor{Q}{_\mu^{AB}}$ is the non-holonomic non-metricity associated to the connection $\tensor{A}{^A_{B\mu}}$. Notice how in non-holonomic frames, it is the vanishing of non-metricity --- rather than torsion --- that is related to symmetries of the connection. Whenever $\tensor{Q}{_\mu^{AB}}=0$, then $\tensor{A}{^{AB}_\mu}=-\tensor{A}{^{BA}_\mu}$. From the symmetry group perspective, this is equivalent to a contraction of $GL(n,\mathbb{R})$ down to one of its (pseudo-)orthogonal subgroups. In our case, $SO\left(1,n-1\right)$. This is, of course, the reason why we blindly did the opposite, in the end of Section~\ref{sec:holonomic_and_non-holonomic_frames}. We expand on that in Section~\ref{sec:geometric_picture}.

\subsection{EC \textit{vs.} ECSK \textit{vs.} GR}\label{ssec:inequivalence}

At first glance, the set of field equations~\eqref{eq:holonomic-field-eqs} seems to deviate from $n$-dimensional Einstein equations. Nonetheless, their physical solutions are still exclusively Einstein $n$-manifolds. This is due to the fact that the EHP action~\eqref{eq:holonomic-EP-action} is explicitly invariant under local projective transformations
\begin{subequations}\label{eq:r-symmetry}
  \begin{align}
    g_{\mu\nu}                          & \rightarrow g_{\mu\nu} \;,                                                              \\
    \tensor{\Gamma}{^\alpha_{\beta\mu}} & \rightarrow \tensor{\Gamma}{^\alpha_{\beta\mu}}+\tensor{\delta}{^\alpha_\beta}U_\mu \;.
  \end{align}
\end{subequations}
This is the so-called $R^d$-symmetry, where $U_\mu$ is an arbitrary vector field~\cite{dadhich2012}. One can show that to choose vanishing $\tensor{Q}{_{\alpha\beta\mu}}$ (and, consequentially, $\tensor{T}{^\alpha_{\beta\mu}}$, via~\eqref{eq:holonomic-cartan-like-field-eqs}) or $\tensor{T}{^\alpha_{\beta\mu}}$ (and, consequentially, $\tensor{Q}{_{\alpha\beta\mu}}$, again, via~\eqref{eq:holonomic-cartan-like-field-eqs}) equates to setting $U_\mu=0$. In other words, these choices are nothing but gauge choices for this projective symmetry. $\tensor{ T }{ ^{ \alpha } _{ \beta \mu } }$ and $ Q_{ \alpha \beta \mu } $ are pure $R^d$-gauge quantities and the traditional EH action --- the action of GR --- is just an $R^d$-gauge fixed version of~\eqref{eq:holonomic-EP-action}. This is true as long as we remain decoupled from matter sources carrying non-vanishing hypermomentum currents. If, for instance, spinorial matter is present, $ \tensor{ T }{ ^{ \alpha } _{ \beta \mu } } $ couples to the spin density tensor and assumes an $R^d$-gauge invariant character. This results in a space of solutions containing Riemann-Cartan $n$-manifolds, and~\eqref{eq:holonomic-EP-action} is an $R^d$-gauge unfixed version of EC theory, not GR\@.

The VEP action~\eqref{eq:vep-action} also enjoys invariance under local projective transformations of the connection, namely,
\begin{subequations}\label{eq:nonholonomic-projective-transformation}
  \begin{align}
    \tensor{e}{^A_\mu}    & \rightarrow \tensor{e}{^A_\mu} \;, \label{eq:projective-transf-vielbein}                                 \\
    \tensor{A}{^A_{B\mu}} & \rightarrow \tensor{A}{^A_{B\mu}}+\tensor{\delta}{^A_B}V_\mu \;, \label{eq:projective-transf-connection}
  \end{align}
\end{subequations}
where $V_\mu$ is an arbitrary vector field. The previous scenario then repeats itself in a non-holonomic fashion. $\tensor{Q}{_\mu^{AB}}$ is a pure $R^d$-gauge quantity proportional to $V_\mu$~\cite{dadhich2012}. In the $R^d$-gauge choice $V_\mu=0$, $\tensor{Q}{_\mu^{AB}}$ and, consequentially, $\tensor{T}{^A_{\mu\nu}}$, via~\eqref{eq:nonholonomic-cartan-like-field-eqs}, vanishes. Again, physical solutions are exclusively Einstein $n$-manifolds as long as there are no couplings to matter sources carrying hypermomentum currents. Otherwise, the solutions are Riemann-Cartan $n$-manifolds and~\eqref{eq:vep-action} is an $ R^{ d } $-gauge unfixed version of ECSK theory.

We just argued that EC and ECSK theory, in non-degenerate spacetime regions and in absence of hypermomentum currents, are both equivalent to GR\@. This was due to the presence of a projective gauge symmetry. Therefore, it is reasonable to expect EC and ECSK to be equivalent to each other \textit{modulus} some gauge artifacts. The non-holonomic to holonomic equivalence is achieved, at the level of field equations and action functionals, by the set of field transformations
\begin{subequations}\label{eq:field-transformation}
  \begin{align}
    g_{\mu\nu}                        & = \tensor{e}{^A_\mu}\tensor{e}{^B_\nu}g_{AB} \;, \label{eq:metrics}                                                                          \\
    \tensor{\Gamma}{^\alpha_{\mu\nu}} & = \tensor{e}{^\alpha_A}\tensor{A}{^A_{B\mu}}\tensor{e}{^B_\nu}-\tensor{e}{^\alpha_A}\partial_\mu\tensor{e}{^A_\nu}\;. \label{eq:connections}
  \end{align}
\end{subequations}
Three points are important to be emphasized about them: (i) the Jacobian matrix of such transformations is clearly not trivial; (ii) it is not even a square matrix; (iii) these transformations assume $\tensor{e}{^\alpha_A}$ exists.

Point (i) is of major importance in the study of this (in)equivalence within a path integral quantization of both theories. It indicates the appearance of non-trivial insertions if one transforms the functional measure from $\mathcal{D}e\mathcal{D}A$ to $\mathcal{D}g\mathcal{D}\Gamma$. Point (ii) reflects the one-to-many nature of the \textquote{inverse} transformations. While a non-holonomic description has a unique holonomic counterpart, a holonomic description has infinitely many non-holonomic versions. All of these versions are $GL\left(n,\mathbb{R}\right)$ gauge transformations of each other --- thus, define only one single physical theory. Finally, point (iii) can be relaxed at the expense of a fixed space topology~\cite{geroch1967,tipler1977,horowitz1991,borde1994,borde1999,heveling2022}, and the equivalence between EC and ECSK\@.

The above facts are well established and can be summarized in the following commutative diagram:

\begin{equation}\label{dia:equivalence}
  \begin{tikzcd}[row sep=huge, column sep=huge]
    \text{\eqref{eq:holonomic-EP-action}} \arrow[d, Rightarrow, "\delta S = 0" description]
    \arrow[r, Leftarrow, "\eqref{eq:field-transformation}"]
    &
    \text{\eqref{eq:vep-action}} \arrow[d, Rightarrow, "\delta S =0" description]
    \\
    \text{\eqref{eq:holonomic-field-eqs}}
    \arrow[r, Leftarrow, "\eqref{eq:field-transformation}"]
    &
    \text{\eqref{eq:nonholonomic-field-eqs}}
  \end{tikzcd} \;. \tag{Diagram 1}
\end{equation}
It is not known in the literature how general this equivalence is. In particular, if it is valid for general metric-affine dynamics --- which lacks projective symmetry. As mentioned in Section~\ref{sec:introduction}, we do expect it to break quantum mechanically. And, classically, we do know~\ref{dia:equivalence} breaks in degenerate spacetime regions. The present work provides a more complete picture of this scenario. In Section~\ref{sec:on-shell_equivalence}, we extend~\ref{dia:equivalence} to very general metric-affine dynamics at any spacetime dimensions. The results make very clear that an invertible \textit{vielbein} is a general requirement for it to hold. Additionally, in Section~\ref{sec:geometric_picture}, we give its geometrical formulation, and clarify its underlying physical meaning.

%% file: content/sections/inequivalence.tex
\section{The general (in)equivalence}\label{sec:on-shell_equivalence}

In order to address the extension of~\ref{dia:equivalence} to the most general metric-affine case, it is convenient to abstract the situation to that of a field theory for the multi-field $\Phi^I$, which single-handedly represents an arbitrary (but finite) number of scalar, vector, 2-tensor, 3-tensor, \dots, $q$-tensor fields\footnote{Here upper-case Latin letters do not simply ranger from $0$ to $n-1$, as it labels the fields within the $\Phi$ multiplet.}
\begin{equation}
  \Phi^I \in \left\{\phi^1,\ldots,\phi^{q_0},\phi_\mu^1, \ldots,\phi^{q_1}_\mu,\phi^1_{\mu\nu},\ldots, \phi^{q_2}_{\mu\nu}, \ldots, \phi^1_{\mu_1\cdots \mu_q},\ldots,\phi^{q_q}_{\mu_1\cdots \mu_q} \right\} \;.
\end{equation}
Its dynamics is encoded in the very general action functional containing up to the $k$-th order derivative of $\Phi^I$,
\begin{equation}
  S\left[\Phi^I\right] = \int d^n x \mathcal{L} \left(\Phi^I, \partial_\mu \Phi^I , \ldots  ,\partial_{\mu_1}\cdots\partial_{\mu_k} \Phi^I  \right)\;.
\end{equation}
Hamilton's principle states that classical configurations of $\Phi^I$ are \textit{extrema} of such functional. Thus, solutions to the partial differential equation
\begin{equation}
  \label{eq:el-equations}
  \delta^{(k)}_{\Phi^I}\mathcal{L}=0 \;,
\end{equation}
known as Euler-Lagrange equation. The notation employed for the Lagrange operator reads
\begin{equation}
  \delta^{(k)}_{\Phi^I} \equiv \sum_{j=0}^{k} \sum_{\mu_1 \leq \cdots \leq \mu_j} {\left( -1 \right)}^j \partial_{\mu_1} \ldots \partial_{\mu_j} \left[ \frac{ \partial \phantom{ \left( \partial_{\mu_1} \ldots \partial_{\mu_j} \Phi^I \right) } }{ \partial \left( \partial_{\mu_1} \ldots \partial_{\mu_j} \Phi^I \right) } \right] \;,
\end{equation}
where the second sum is over the ordered set $\left\{ \left( \mu_1,\ldots,\mu_j \right) \; ; \; \mu_1 \leq \cdots \leq \mu_j \right\}$. Whenever this set is empty, which is the case for $j=0$, the operator $\partial_{\mu_1} \ldots \partial_{\mu_j}$ is the identity --- we end up with just a $\partial/\partial \Phi^I$ contribution.

It is a common exercise in field theory to consider the field transformations
\begin{equation}
  \label{eq:standard-field-transf}
  \Phi'^{J}=\Phi'^{J} \left(\Phi^I\right) \;,
\end{equation}
whose Jacobian matrix
\begin{equation}
  \tensor{J}{^{J}_I} \equiv \frac{\partial \Phi'^{J}}{\partial \Phi^I}
\end{equation}
is square and non-singular. As a result, Euler-Lagrange field equations~\eqref{eq:el-equations} transform covariantly,
\begin{equation}
  \label{eq:el-equations-covariant-transf}
  \tensor{J}{^J_I}\delta^{(k)}_{\Phi'^{J}}\mathcal{L}'\left(\Phi'^{J}, \partial_\mu \Phi'^{J}, \ldots , \partial_{\mu_1\cdots\mu_k} \Phi'^{J}\right) = 0 \;.
\end{equation}
Since $J^{-1}$ exists, it can be used in~\eqref{eq:el-equations-covariant-transf} to yield
\begin{equation}
  \label{eq:el-equation-prime}
  \delta^{(k)}_{\Phi'^{J}}\mathcal{L}'=0 \;.
\end{equation}
This establishes the on-shell equivalence between $\Phi^I$ and $\Phi'^{J}$ theories\footnote{This latter step is what fails in degenerate spacetime regions.}.

On the other hand, it is less standard to consider field transformations of the form
\begin{equation}
  \label{eq:non-standard-field-transf}
  \Phi'^{J} = \Phi'^{J} \left(\Phi^I, \partial_\mu \Phi^I\right) \;.
\end{equation}
Its non-vanishing dependence on first order derivatives w.r.t.\ the fields
\begin{equation}
  \label{eq:Q}
  \tensor{K}{^J_I^\mu}\equiv \frac{\partial \Phi'^{J}}{\partial \left(\partial_\mu \Phi^I\right)} \;
\end{equation}
results in the Jacobian matrix
\begin{equation}
  \label{eq:non-standard-jacobian}
  \mathbb{J}^\mu = \begin{bmatrix}
    \tensor{J}{^{J}_I}\tensor{\delta}{_0^\mu} \\
    \tensor{K}{^J_I^\mu}
  \end{bmatrix}
  \;
\end{equation}
having twice as many rows than columns. This is a telltale sign of singular, non-invertible transformations. More precisely, if one were to invert~\eqref{eq:non-standard-field-transf}, one would quickly realize that the system of equations that needs to be solved is undetermined, having infinitely many solutions. Thus, this is precisely the abstraction of the gravitational case presented in the end of Section~\ref{sec:gravities}, equation~\eqref{eq:field-transformation}. It is a tediously long, but straightforward, calculation to show that, under~\eqref{eq:non-standard-field-transf}, Euler-Lagrange field equations~\eqref{eq:el-equations} transform according to
\begin{equation}
  \label{eq:non-standard-field-eq-transf}
  \tensor{J}{^{J}_I}\delta^{(k)}_{\Phi'^{J}}\mathcal{L}'=\partial_\mu \left[\tensor{K}{^{J}_I^\mu}\delta^{(k)}_{\Phi'^{J}}\mathcal{L}'\right] \;.
\end{equation}
To the author knowledge, equation~\eqref{eq:non-standard-field-eq-transf} is not present in the literature and clearly a non-covariant behavior. At this point, $\Phi^I$ and $\Phi'^{J}$ theories have no chance to be equivalent, even if $J^{-1}$ were at our disposal.

In order to regain some sense of covariance, the right-hand side of~\eqref{eq:non-standard-field-eq-transf} has to vanish --- $\tensor{K}{^{J}_I^\mu}\delta^{\left(k\right)}_{\Phi'^{J}}\mathcal{L}'$ has to be divergence-free. This is trivially achieved in the case of $\tensor{K}{^{J}_I^\mu}=0$, which corresponds to the canonical field transformations~\eqref{eq:standard-field-transf}. The gravitational case, on the other hand, has a non-vanishing $\tensor{K}{^1_2^\mu}$. At first, this raises the suspicion that $\Phi^I$ gravity theories cannot possibly be equivalent to $\Phi'^J$ ones.

Let us take $\Phi^I \in \left\{\tensor{e}{^A_\mu}, \tensor{A}{^A_{B\mu}}\right\}$, ${\Phi'}^J \in \left\{g_{\mu\nu}, \tensor{\Gamma}{^\alpha_{\beta\mu}} \right\}$ and the field transformations of the kind~\eqref{eq:non-standard-field-transf} to be~\eqref{eq:field-transformation}. This yields the Jacobian matrices
\begin{equation}
  \label{eq:j-jacobian-for-gravity}
  \tensor{J}{^{J}_I} = \begin{bmatrix}
    \frac{\partial g_{\mu\nu}}{\partial \tensor{e}{^A_\gamma}} & \frac{\partial \tensor{\Gamma}{^\alpha_{\beta\mu}}}{\partial \tensor{e}{^A_\gamma}}    \\
    0                                                          & \frac{\partial \tensor{\Gamma}{^\alpha_{\beta\mu}}}{\partial \tensor{A}{^A_{B\gamma}}}
  \end{bmatrix} \;,
\end{equation}
and
\begin{equation}
  \label{eq:k-jacobian-for-gravity}
  \tensor{K}{^J_I^\lambda} = \begin{bmatrix}
    0 & \frac{\partial \tensor{\Gamma}{^\alpha_{\beta\mu}}}{\partial \left(\partial_\lambda\tensor{e}{^A_\gamma}\right)} \\
    0 & 0
  \end{bmatrix} \;.
\end{equation}
Thus, the transformed Euler-Lagrange field equations~\eqref{eq:non-standard-field-eq-transf} reduce to
\begin{subequations}\label{eq:non-standard-field-eq-transf-for-gravity}
  \begin{align}
    \left(\frac{\partial g_{\mu\nu}}{\partial \tensor{e}{^A_\gamma}} \delta^{(k)}_{g_{\mu\nu}} + \frac{\partial \tensor{\Gamma}{^\alpha_{\beta\mu}}}{\partial \tensor{e}{^A_\gamma}} \delta^{(k)}_{\tensor{\Gamma}{^\alpha_{\beta\mu}}}\right) \mathcal{L}' & = \partial_\lambda \left(\frac{\partial \tensor{\Gamma}{^\alpha_{\beta\mu}}}{\partial \left(\partial_\lambda \tensor{e}{^A_\gamma}\right)} \delta^{(k)}_{\tensor{\Gamma}{^\alpha_{\beta\mu}}}\mathcal{L}' \right) \;, \label{eq:transformed-einstein-like-field-eq} \\
    \frac{\partial \tensor{\Gamma}{^\alpha_{\beta\mu}}}{\partial \tensor{A}{^A_{B\gamma}}} \delta^{(k)}_{\tensor{\Gamma}{^\alpha_{\beta\mu}}} \mathcal{L}'                                                                                                  & = 0 \;. \label{eq:transformed-cartan-like-field}
  \end{align}
\end{subequations}
Individually, the transformed Cartan-like field equation~\eqref{eq:transformed-cartan-like-field} does behave covariantly due to the vanishing of $\tensor{J}{^2_1}$, $\tensor{K}{^2_1^\lambda}$ and $\tensor{K}{^2_2^\lambda}$. The transformed Einstein-like field equation~\eqref{eq:transformed-einstein-like-field-eq} does not. Collectively, \eqref{eq:transformed-cartan-like-field}~and~\eqref{eq:transformed-einstein-like-field-eq} do form a system. And, a solution for~\eqref{eq:transformed-cartan-like-field} has to be a solution for~\eqref{eq:transformed-einstein-like-field-eq}. From~\eqref{eq:transformed-cartan-like-field}, it is clear that $\delta^{(k)}_{\tensor{\Gamma}{^\alpha_{\beta\mu}}}\mathcal{L}'=0$ if $\partial \tensor{\Gamma}{^\alpha_{\beta\mu}} / \partial \tensor{A}{^A_{B\gamma}}$ is an invertible matrix. If so, $\delta^{(k)}_{\tensor{\Gamma}{^\alpha_{\beta\mu}}}\mathcal{L}'$ also vanishes in~\eqref{eq:transformed-einstein-like-field-eq}, thereby killing all undesirable terms. Thus,~\ref{dia:equivalence} can be extended to very general metric-affine dynamics.

In conclusion, if we consider
\begin{subequations}
  \begin{align}
    \delta^{(k)}_{\tensor{e}{^A_\gamma}}\mathcal{L}     & = 0 \;, \\
    \delta^{(k)}_{\tensor{A}{^A_{B_\gamma}}}\mathcal{L} & = 0 \;,
  \end{align}
\end{subequations}
for whichever chosen $\mathcal{L}$, and apply field transformations~\eqref{eq:field-transformation}, we end up with
\begin{subequations}\label{eq:non-standard-field-eq-transf-for-gravity-simplified}
  \begin{align}
    \delta^{(k)}_{g_{\mu\nu}} \mathcal{L}'                          & = 0 \;, \\
    \delta^{(k)}_{\tensor{\Gamma}{^\alpha_{\beta\mu}}} \mathcal{L}' & = 0 \;,
  \end{align}
\end{subequations}
as long as $\partial \tensor{\Gamma}{^\alpha_{\beta\mu}} / \partial \tensor{A}{^A_{B\gamma}}$ and $ \partial g_{ \mu \nu } / \partial \tensor{e}{^A_{\gamma}} $ (both related to $\tensor{e}{^B_\mu}$) are non-singular. Thus, gravity theories, formulated in holonomic \textit{versus} non-holonomic frames, are on-shell equivalent in a way that is independent of the particular metric-affine dynamics and/or spacetime dimension. This result is largely due to the functional form of field transformations~\eqref{eq:field-transformation} and the invertible \textit{vielbein} condition.

%% file: content/sections/bundles.tex
\section{The geometrical framework}\label{sec:geometric_picture}

In other to explain the functional form~\eqref{eq:field-transformation}, and clarify its physical meaning, we need to investigate the fiber bundle structures over $X$ --- implicitly used in both the holonomic and non-holonomic frame descriptions of gravity. We formalize holonomic quantities in terms of natural bundles, and non-holonomic ones in terms of soldered bundles.

\subsection{Natural bundles}\label{ssec:natural_bundles}

Consider the set $TX \equiv \left\{ \sqcup_x T_x X \; \forall \; x\in X \right\}$, and the map $\pi_{TX}: TX \rightarrow X$. $TX$ is called the total space\footnote{Sometimes we might use the total space to refer to the whole bundle structure.}, and it inherits a $2n$-manifold structure from $X$; $\pi_{TX}$ is called the projection map, and it is smooth and surjective. The typical fiber, $ { \pi_{TX} }^{-1} \left( x \right) $, is isomorphic to $ T_x X $ and, as such, carries a $ GL \left( n, \mathbb{R} \right)$ representation reminiscent of smooth changes of coordinates --- mentioned in Section~\ref{sec:holonomic_and_non-holonomic_frames}. This later statement is the archetypical examples of a categorical lift. The structure just described is known as the tangent bundle of $X$.

Morphisms in the category of smooth manifolds induce morphisms in the category of smooth vector bundles via a functor $ \mathcal{F} $~\cite{tu2011}. This can be neatly captured by the commutative diagram
\begin{equation}
  \label{dia:categorical_lift}
  \begin{tikzcd}
    \mathcal{F}X
    \arrow[r, "\mathcal{F}f"]
    \arrow[d, "\pi_{\mathcal{F}X}" left]
    &
    \mathcal{F}X'
    \arrow[d, "\pi_{\mathcal{F}X'}" ]
    \\
    X
    \arrow[r, "f"]
    &
    X'
  \end{tikzcd} \tag{Diagram 2} \;.
\end{equation}
The bundle morphism $ \mathcal{F} f $ is said to be the functorial lift of the morphism $ f $. Bundles above $ X $ constructed in this functorial way are said to be natural~\cite{kolar2000}. The tangent bundle is the natural bundle obtained by considering $ X' = X$, $ f $ as a local automorphism\footnote{Defined as map from $ X $ to $ X $ which is necessarily a diffeomorphism only on a chart, \textit{i.e.}, $ f $ is not necessary a diffeomorphism but $ f|_U: U \rightarrow f(U) $ is.} and $\mathcal{F}$ as the tangent (pushforward) map $T$. On $U$, this translates to transition functions $ x^{\nu'} \left( x^\mu \right) $ lifting to automorphisms $\tensor{J}{^\mu_{\nu'}}(x)$ on $T_x U$. Since the group of automorphisms of the typical fiber is isomorphic to the structure group of the bundle itself, the tangent bundle is constrained to have $ GL \left( n, \mathbb{R} \right) $. In other words, $ lAut \left( X \right) $ functorially lifts to $ TX $ as $ GL\left( n,\mathbb{R} \right) $. Other natural bundles on $ X $ can be defined by only changing the typical fiber to another representation space of $ GL \left( n, \mathbb{R} \right) $. The co-tangent bundle $T^* X$ is the one with typical fiber isomorphic to ${T}^*_x X$\footnote{Categorically, $\mathcal{F}$ is the co-functor $T^*$ of $T$.} and, more generally, the $(r,s)$-tensor bundle $\mathcal{T}^s_r X$ is the one with typical fiber isomorphic to $T_x X^{\otimes^r}\otimes T^*_x X^{\otimes^s}$\footnote{Categorically, $\mathcal{F}$ is the tensor products of functors $\mathcal{T}^s_r\equiv T^{ \otimes^r } \otimes {T^*}^{ \otimes^s }$.}.

A right inverse for $\pi_{TX}$ is a tangent vector field on $X$. It is called a section of this bundle, and defined as a map $\sigma_{TX}: U \subseteq X \rightarrow TX$ such that $\pi_{TX} \circ \sigma_{TX} = \mathds{1}_U$. There might be topological obstructions for the set equality to hold for a nowhere vanishing $\sigma_{TX}$. Most nowhere vanishing sections are local ($U\subset X$). Global ones ($U=X$) are only guaranteed to exist in trivial bundles --- $TX$ needs to be globally diffeomorphic to $X \times \mathbb{R}^n$. The canonical example is given by the hairy ball theorem. No nowhere vanishing tangent vector field globally exists on the $S^2$. $TS^2$ is not diffeomorphic to $S^{2} \times \mathbb{R}^{2}$. The opposite is true for $S^3$. $TS^{3} \cong S^{3} \times \mathbb{R}^{3}$, and it is one reason why it can accept an $SU(2)$ Lie group structure.

The above results are closely connected to the value of their Euler characteristic: $\chi \left( S^{2} \right) = 2$ and $ \chi \left( S^{3} \right) = 0$. In such topologies, $\chi \left( X \right)$ acts as the obstruction for the existence of any nowhere vanishing global section in $TX$. It so happens that $\chi \left( S^n \right)=0$, if $n$ is odd; and $\chi \left( S^n \right)=2$, if $n$ is even. In the former case, $n \in \left\{ 1, 3, 7 \right\}$ are especial since they are the only ones with $TS^n \cong S^n \times \mathbb{R}^n$.

The space of all sections in $TX$ is denoted as $\Gamma \left( TX \right)$. One can infer now that the metric field $ g(x) $ is an element in the symmetric subspace $ \Gamma ( \text{\raisebox{\depth}{\scalebox{1}[-1]{$\Lambda$}}}^2 X ) \subset \Gamma ( \mathcal{T}^2_0 X )$ --- and, depending on the topology of $X$, $ \chi \left( X \right)$ might act as an obstruction for it to be a nowhere vanishing globally defined field.

A very important natural bundle is the frame bundle $\pi: FX \rightarrow X$ of $TX$. The structure group and base space are the same as before, but the total space $FX$ is defined as the set of all tangent frames on $X$. In particular, the typical fiber $F_x X$ is diffeomorphic to $GL\left(n, \mathbb{R}\right)$ itself. This makes $FX$ $n\left(n+1\right)$-dimensional. Furthermore, $F_x X$ carries a smooth and free right action of $GL\left(n, \mathbb{R}\right)$. These facts make $ FX $ into a $GL\left(n, \mathbb{R}\right)$ principal bundle over $X$. This is the space where holonomic frames live in. In particular, $\partial_\mu|_x$ is an element of $F_x X$ and the holonomic moving frame $\partial_\mu \left(x\right)$ is an element of $\Gamma\left(FX\right)$. Finally, transformation laws~\eqref{eq:holonomic-frame-transf-rule} and $\eqref{eq:holonomic-coframe-transf-rule}$ are just a reflex of the naturalness of this bundle, \textit{i.e.}, that $f$ canonically lifts to it and to the co-frame bundle $F^* X$, respectively.

$FX$ has such importance because all other natural vector bundles over $X$ can be derived from it via the associated vector bundle construction. Let $\rho: GL\left(n, \mathbb{R}\right) \rightarrow GL\left(\mathcal{V}\right)$ be a representation of $GL\left(n, \mathbb{R}\right)$ on a vector space $\mathcal{V}$. One can show that the product space $FX \times \mathcal{V}$, \textit{modulus} the equivalence relation $ \left(u, v\right) \sim \left(u g, \rho\left(g^{-1}\right)v\right)$, where $ u \in FX $, $v \in\mathcal{V}$, and $ g\in GL\left(n, \mathbb{R}\right)$, does form a vector bundle over $X$. This bundle, $\pi_{ A \left( \mathcal{V} \right) }: A\left(\mathcal{V}\right)\rightarrow X$, where $A\left(\mathcal{V}\right)\equiv FX \times \mathcal{V}/\sim$, is said to be a vector bundle associated to $FX$. Clearly, there are as many associated vector bundles to $FX$ as there are representation spaces of $GL\left(n, \mathbb{R}\right)$. In particular, $ A \left( T_x X\right ) $ is an associated vector bundle trivially isomorphic to $TX$, $ A \left( {T}^*_x X \right) $ is to $T^*X$, and so on and so forth. In this sense, all natural vector bundles are derived from $FX$ --- they are natural because $ FX $ is.

If $X$ is paracompact, $FX$ can have its own tangent bundle decomposed into vertical and horizontal sub-bundles, $ TFX = T_V FX \oplus T_H FX $. While $T_V FX$ is uniquely defined as the kernel of $\pi_*$, its complement $T_H FX$ is not. Given a $\rho$-equivariant\footnote{A $\mathcal{V}$-valued form $\phi$ on $FX$ is $\rho$-equivariant if, for every $ g \in GL \left( n, \mathbb{R} \right) $, \[ {R}^{*}_{g} \left( \phi \right) = \rho \left( {g}^{-1} \right) \phi \;, \] where $ {R}^{*}_{g} $ is the pullback via the right action $ {R}_{g} $ of $ GL \left( n,\mathbb{R} \right) $ on $ FX $.} $ \mathfrak{gl} \left( n, \mathbb{R} \right) $-valued global section $\omega$ in $\Gamma\left({T}_{V}^{*} FX\right)$, which is the identity on $ \Gamma \left ( T_V FX \right) $, the choice $ \Gamma \left( T_H FX \right) = \ker \left( \omega \right)$ can be made. $\omega$ is a connection form on $FX$. This construction gives a recipe on how to differentiate $\rho$-equivariant $\mathcal{V}$-valued $k$-forms into $\rho$-equivariant $\mathcal{V}$-valued $\left(k+1\right)$-forms on $FX$. $\omega$ is itself an example of such forms. However, it is vertical\footnote{Vertical means it annihilates sections in $ \Gamma \left( T_H FX \right) $.}, which means it differentiates itself in an unusual way, resulting in
\begin{equation}\label{eq:curvatureform}
  \Omega\equiv d\omega+\frac{1}{2}\left[\omega,\omega\right] \;,
\end{equation}
where $d$ is the exterior derivative and $\left[\;,\;\right]$ is the graded Lie bracket. $\Omega$ is a $\rho$-equivariant $\mathfrak{gl}\left(n,\mathbb{R}\right)$-valued 2-form: the curvature form of $\omega$. This process of differentiation abhors verticality. $\Omega$ is horizontal\footnote{Horizontal means it annihilates sections in $ \Gamma \left( T_V FX \right) $.}, and so is the result of every differentiation via $\omega$. Thus, such procedure is better understood as an operation on the space $\Gamma\left(\Lambda_{H,\rho}^*FX\right)$ of horizontal $\rho$-equivariant $\mathcal{V}$-valued forms on $FX$. Let $\rho_*|_{\mathds{1}}: \mathfrak{gl}\left(n,\mathbb{R}\right) \rightarrow \mathfrak{gl}\left(\mathcal{V}\right)$ be pushforward map via $\rho$ at the identity element $\mathds{1}$ in $GL\left(n,\mathbb{R}\right)$. Then, $\omega$ indeed defines an endomorphism
\begin{equation}
  \label{eq:ext.cov.der.}
  D=d+\rho_*|_{\mathds{1}}\left(\omega\right) \;
\end{equation}
on $\Gamma\left(\Lambda^*_{H,\rho} FX\right)$ that maps $\Gamma\left(\Lambda^k_{H,\rho} FX\right)$ into $\Gamma(\Lambda^{k+1}_{H,\rho} FX)$ while satisfying the graded Leibniz rule. This is an exterior covariant derivative on $FX$.

The space $\Gamma\left(\Lambda^*_{H,\rho}FX\right)$ plays a pivotal role since an isomorphism exists between it and the space $\Gamma\left(A\left(\mathcal{V}\right)\otimes \Lambda^*X\right)$ of $\mathcal{V}$-valued forms on $X$. Using such map, $D$ descends from $FX$ to each $A\left(\mathcal{V}\right)$ as an operator $D_{A\left(\mathcal{V}\right)}$ that, instead, differentiates elements in $\Gamma\left(A\left(\mathcal{V}\right)\otimes \Lambda^k X \right)$ into elements in $\Gamma\left(A\left(\mathcal{V}\right)\otimes \Lambda^{k+1}X\right)$. Now, one is able to guess that $\nabla$, introduced in Section~\ref{sec:gravities}, is just $D_{TX}$ composed with the interior product $\rfloor$ in $\Gamma\left(\Lambda^*X\right)$,
\begin{equation}
  \label{eq:cov.diff.}
  \nabla \equiv \;\rfloor D_{TX} \;.
\end{equation}
This properly sends elements from $\Gamma\left(TX\otimes TX\right)$ to $\Gamma\left(TX\right)$.

$ \Omega \in \Gamma\left(\mathfrak{gl}\left( n, \mathbb{R} \right) \otimes \Lambda^2_{H,\rho}FX\right)$ descends to an element $R \in \Gamma\left(\mathfrak{gl}\left(n,\mathbb{R}\right)\otimes\Lambda^2X\right)$. $ R $ is the familiar $\mathfrak{gl}\left(n,\mathbb{R}\right)$-valued curvature 2-form on $X$ --- the geometrical structure behind $\tensor{R}{^\alpha_{\beta\mu\nu}}$. On the other hand, $\omega$ cannot descend to $X$ via the associated bundle construct since it is a vertical form. Nevertheless, one can always use a section $\sigma_{FX}: U \subseteq X \rightarrow FX$ to pull it down from $\Gamma\left(\Lambda^1_{V,\rho}FX\right)$ to $\Gamma\left(\mathfrak{gl}\left(n,\mathbb{R}\right)\otimes \Lambda^1 U\right)$,
\begin{equation}
  \label{eq:local_connection}
  \Gamma \equiv {\sigma_{FX}}^* \omega \;,
\end{equation}
where ${\sigma_{FX}}^*$ is the pullback map. $\Gamma$ is the $\mathfrak{gl}\left(n,\mathbb{R}\right)$-valued 1-form on $U$ which we would call as a $GL\left(n,\mathbb{R}\right)$ gauge field in the traditional sense --- the geometrical structure behind the affine connection $\tensor{\Gamma}{^\alpha_{\beta\mu}}$.

As one can see, gravity is described by a peculiar kind of gauge theory, in the sense that the fundamental fields capture the dynamics of the base space $X$ itself. The holonomic way accomplishes this by defining the theory directly on natural bundles. After all, these are the bundles having, by definition, functorial lifts of $lAut\left(X\right)$ --- thus, a direct connection with $X$. However, this is not the only way to do it.

\subsection{Soldered bundles}\label{ssec:soldered_bundles}

Consider that $X$ has such topology that, given a manifold $P$ and a Lie group $G$, the non-trivial principal $G$-bundle $\pi': P \rightarrow X$ also exists over it. Moreover, that there exists the map $h: FX \rightarrow P$ such that $\pi=\pi'\circ h$. Again, this principal bundle morphism can be neatly captured by the commutative diagram
\begin{equation}
  \label{dia:g-bundle_isomorphism}
  \begin{tikzcd}
    FX
    \arrow[r,"h"]
    \arrow[d,"\pi" left]
    &
    P
    \arrow[d, "\pi'" ]
    \\
    X
    \arrow[r, "\mathds{1}_X"]
    &
    X
  \end{tikzcd} \tag{Diagram 3} \;,
\end{equation}
where $\mathds{1}_X$ is the identity automorphism on $X$. $h$ is called vertical since it covers $\mathds{1}_X$.

It is important to note that, at each fiber $\pi^{-1}(x)$, $h$ defines a homomorphism of Lie groups $h|_{\pi^{-1}}:GL\left(n,\mathbb{R}\right)\rightarrow G$. Whenever $h|_{\pi^{-1}}$ is the actual identity automorphism on $GL\left(n,\mathbb{R}\right)$, we call $h$ equivariant. If this is the case, $P$, in~\ref{dia:g-bundle_isomorphism}, is said to be soldered to $X$. Let us assume so, and that $\omega'$ is a connection on $P$. This connection also labeled as soldered since
\begin{equation}
  \label{eq:soldered_connection}
  \omega = h^* \omega' \;,
\end{equation}
where $h^*$ is the pullback map via $h$.

It is a trivial fact that $FX$ is soldered to itself via vertical equivariant automorphisms. It corresponds to the case where $P=FX$. In such scenario, equation~\eqref{eq:soldered_connection} represents a gauge transformation on $FX$. Indeed, the set of all vertical equivariant automorphisms on $FX$, denoted as $\mathcal{G}\left(FX\right)$, is the set of all gauge transformations on $FX$~\cite{bleecker1981, rudolph2017}.

Moving on, consider a representation $\rho': G\rightarrow GL\left(\mathcal{V}'\right)$ of $G$ on $\mathcal{V}'$. Exclusively on soldered $G$-bundles, there exists an element $\theta \in \Gamma \left(\Lambda^1_{H,\rho'}P\right)$ such that $\dim \left(\mathcal{V}'\right)=n$. Let $D'$ be the exterior covariant derivative associated with $\omega'$, the so-called torsion form $\Theta \in \Gamma \left(\Lambda^2_{H,\rho'}P\right)$ can be defined as
\begin{equation}
  \label{eq:torsion_form}
  \Theta\equiv D'\theta \;.
\end{equation}

Via the associated vector bundle construction regarding $\rho'$, $\theta$ as well as $\Theta$ descend to $A'\left(\mathcal{V}'\right)$, respectively, as an element $e\in \Gamma \left(A'\left(\mathcal{V}'\right)\otimes \Lambda^1X\right)$, which can be regarded as a vertical vector bundle isomorphism,
\begin{equation}
  \label{dia:v-bundle_isomorphism}
  \begin{tikzcd}
    TX
    \arrow[r,"e"]
    \arrow[d,"\pi_{TX}" left]
    &
    A'\left(\mathcal{V}'\right)
    \arrow[d, "\pi_{A'\left(\mathcal{V'}\right)}" ]
    \\
    X
    \arrow[r, "\mathds{1}_X"]
    &
    X
  \end{tikzcd} \;, \tag{Diagram 4}
\end{equation}
and an element $T\in \Gamma \left(A'\left(\mathcal{V}'\right)\otimes \Lambda^2X\right)$, given by
\begin{equation}
  \label{eq:torsion_2form}
  T=D_{A'\left(\mathcal{V}'\right)}e \;.
\end{equation}

$e$ is the well-known $\mathcal{V}'$-valued soldering 1-form --- the geometrical quantity behind $\tensor{e}{^A_\mu}$. $T$ is the $\mathcal{V}'$-valued torsion 2-form on $X$ --- the geometrical quantity behind $\tensor{T}{^A_{\mu\nu}}$. These, of course, lack in the traditional (unsoldered) gauge-theoretical framework of particle physics.

As a consistency check, consider, again, $P=FX$. Moreover, consider $FX$ to be trivially soldered, \textit{i.e.}, $\mathcal{G}\left(FX\right)$ contains only the trivial gauge transformation $\omega=\mathds{1}_{FX}^* \omega'$. To fulfill condition $\dim \left(\mathcal{V'}\right)=n$ for $\theta$, let $\mathcal{V}'\simeq T_x X$. In such case, $e$, of course, only corresponds to the vertical identity transformation $\mathds{1}_{TX}$ on $TX$. This tautology implies that $T$ reduces to
\begin{align}
  \label{}
  T & = D_{TX}\mathds{1}_{TX} \;, \nonumber                                       \\
    & = D_{TX}\left(\partial_\alpha\otimes dx^\alpha\right) \;, \nonumber         \\
    & = \partial_\alpha\otimes \left(D_{TX}dx^\alpha\right) \;, \nonumber         \\
    & = \partial_\alpha\otimes dx^\beta \wedge \tensor{\Gamma}{^\alpha_\beta} \;,
\end{align}
where the definition $dx^\beta \wedge \tensor{\Gamma}{^\alpha_\beta} \equiv \rho_*|_{\mathds{1}}\left(\omega\right) dx^\alpha$, in which $\wedge$ is the wedge product, was used. Then,
\begin{align}
  \label{eq:torsion_2form_trivial_solder}
  T\left(\partial_\mu,\partial_\nu\right) & = \partial_\alpha \otimes dx^\beta\wedge\tensor{\Gamma}{^\alpha_\beta} \left(\partial_\mu,\partial_\nu\right) \;,\nonumber                                                          \\
                                          & = \partial_\alpha \otimes \left(\tensor{\delta}{^\beta_\mu}\tensor{\Gamma}{^\alpha_{\beta\nu}} - \tensor{\delta}{^\beta_\nu}\tensor{\Gamma}{^\alpha_{\beta\mu}}\right) \;,\nonumber \\
                                          & = \partial_\alpha \otimes \left(\tensor{\Gamma}{^\alpha_{\mu\nu}} - \tensor{\Gamma}{^\alpha_{\nu\mu}}\right) \;,
\end{align}
which is in agreement with the definition in~\eqref{eq:vanishing_torsion}. One can say that the torsion tensor collapses to the antisymmetric sector of $\Gamma_{\mu\nu}$ once $FX$ is trivially soldered to itself --- which is the case for holonomic theories of gravity.

Finally, consider $ P=FM $, where $ FM $ is the frame bundle of the $ n $-dimensional Minkowski space $ M $. It is constructed over $M$ in the same way $FX$ is constructed over $X$. Thus, $G$ is forced to equal $GL\left(n,\mathbb{R}\right)$. From the perspective of $X$, elements in $FM$, in its associated vector bundles, $A'\left(\mathcal V'\right)$, and the $GL\left(n,\mathbb{R}\right)$ actions over them, are all non-holonomic in nature. We discussed this first in Section~\ref{sec:holonomic_and_non-holonomic_frames}. In the language developed here, this means that these are not natural bundles over $X$.

The non-holonomic frame $\tau_A|_x$, also introduced in Section~\ref{sec:holonomic_and_non-holonomic_frames}, can be regarded as an element of $FM$ over $X$. The moving frame $\tau_A\left(x\right)$ is an element of $\Gamma\left(FM\right)$ over $X$. The non-holonomic $GL\left(n,\mathbb{R}\right)$ connection form $\omega'$ has local projection $A$ and associated curvature $F$ --- the geometrical quantities behind $\tensor{A}{^A_{B\mu}}$ and $\tensor{F}{^A_{B\mu\nu}}$, respectively. These are the gauge-theoretical fields used in the ECSK theory of gravity and its generalizations, \textit{vide}~\eqref{eq:nonholcurvature}\footnote{Although $A$ is a flat connection ($F=0$) if $FM$ is seen as a bundle over $M$, the same is not necessarily true if $FM$ is seen as a bundle over $X$.}. The vector space $V_x$ that $\tau_A|_x$ spans is a fiber of $A'\left(\mathcal{V}'\right)$ over $x$ such that $\dim\left(\mathcal{V}'\right)=n$. In fact, $V\equiv \sqcup_x V_x$ is exactly the kind of vector bundle present in~\ref{dia:v-bundle_isomorphism}. Clearly, $V_x$ is just $T_x M$, and $V$ is just $TM$.

In order to clarify how $e$ solders non-holonomic frames on $X$, consider $\tau^A\in \Gamma\left(T^*M\right)$ and the map $e^*:T^*M \rightarrow T^*X$ where
\begin{equation}
  \label{eq:soldering_pullback}
  \left[e^*\left(\tau^A\right)\right]\left(\partial_\mu\right) = \tau^A\left[e\left(\partial_\mu\right)\right] \;.
\end{equation}
Notice that $e$ maps holonomic frames into non-holonomic ones while $e^*$ glues non-holonomic co-frames on $X$. In practice, due to the contravariant nature of the pullback map, the roles of $\tensor{e}{^A_\mu}$ and $\tensor{e}{^*_\mu^A}$ get flipped. Indeed,
\begin{align}
  \label{eq:holonomic-to-nonholonomic-frame}
  e \left(\partial_\mu\right) & = \tensor{e}{^A_\mu}\tau_A \;,\nonumber \\
                              & = \tensor{e}{^*_\mu^A}\tau_A \;,
\end{align}
while
\begin{align}
  \label{eq:nonholonomic-to-holonomic-coframe}
  e^* \left(\tau^A\right) & = \tensor{e}{^*_\mu^A}dx^\mu \;,\nonumber \\
                          & = \tensor{e}{^A_\mu}dx^\mu \;.
\end{align}
where, as already mentioned, the vielbein field $\tensor{e}{^A_\mu}\equiv \tau^A\left[e\left(\partial_\mu\right)\right]$ is the matrix representation of $e$ in the basis $\tau_A\otimes dx^\mu$. And, $\tensor{e}{^*_\mu^A}\equiv\left[e^*\left(\tau^A\right)\right]\left(\partial_\mu\right)$ is the matrix representation of $e^*$ in $dx^\mu\otimes\tau_A$.

Equation~\eqref{eq:soldering_pullback}, stating that $\tensor{e}{^*_\mu^A}=\tensor{e}{^A_\mu}$, was used in both~\eqref{eq:holonomic-to-nonholonomic-frame} and~\eqref{eq:nonholonomic-to-holonomic-coframe}. In the literature, $e^*\left(\tau^A\right)$ is presented as the 1-form vielbein $e^A$ while $e\left(\partial_\mu\right)$ is mostly ignored. The former is the \textquote{subtle} relation between $e^A$ and $\tau^A$ mentioned in the end of Section~\ref{sec:holonomic_and_non-holonomic_frames}. Additionally, as long as $e$ is an isomorphism, inverses exist for it and its pullback. Explicitly, $\tensor{e}{^\mu_A}\equiv dx^\mu\left[e^{-1}\left(\tau_A\right)\right]$ and $\tensor{e}{^*_A^\mu}\equiv \left[e^{*-1}\left(dx^\mu\right)\right]\left(\tau_A\right)$. Moreover,
\begin{align}
  \label{eq:nonholonomic-to-holonomic-frame}
  e^{-1} \left(\tau_A\right) & = \tensor{e}{^\mu_A}\partial_\mu \;,\nonumber \\
                             & = \tensor{e}{^*_A^\mu}\partial_\mu \;,
\end{align}
while
\begin{align}
  \label{eq:holonomic-to-nonholonomic-coframe}
  e^{*-1} \left(dx^\mu\right) & = \tensor{e}{^*_A^\mu}\tau^A \;,\nonumber \\
                              & = \tensor{e}{^\mu_A}\tau^A \;.
\end{align}
The analog of equation~\eqref{eq:soldering_pullback} for $e^{-1}$ states that $\tensor{e}{^*_A^\mu}=\tensor{e}{^\mu_A}$ and was used in both~\eqref{eq:nonholonomic-to-holonomic-frame} and~\eqref{eq:holonomic-to-nonholonomic-coframe}. It is easy to show that compositions $e^{*-1}\circ e^*$ and $e^{-1}\circ e$ are behind equations~\eqref{eq:inversepullbackvielbein} and~\eqref{eq:inversevielbein}, respectively. In literature, however, it is commonplace to define $e^{-1}\left(\tau_A\right)$ as the 1-vector $e_A$ such that $e^A \left(e_B\right)=\tensor{\delta}{^A_B}$. This latter equation does not hold by itself, but it is a consequence of~\eqref{eq:inversepullbackvielbein} being true.

\subsection{The geometric equivalence principle}\label{ssec:the_equivalence_principle}

The last geometrical structure we need to address is that of a metric. In the beginning of this section, we commented on how a metric tensor on $X$ is an element of $\Gamma (\text{\raisebox{\depth}{\scalebox{1}[-1]{$\Lambda$}}}^2 X)$. Such a metric lives on a natural bundle and thus is holonomic in nature. Analogous definition can be made on using any other vector bundle over $X$. For instance, a non-holonomic metric $g'$ on $X$ can be defined as an element living in $\Gamma (\text{\raisebox{\depth}{\scalebox{1}[-1]{$\Lambda$}}}^2A'\left(\mathcal{V}'\right))$.

We also made comments on how there might be topological obstructions for a local section to be smoothly glued together to form a global one. On certain topologies, $\chi\left(X\right)$ plays that role. Luckily, if $X$ is paracompact, partition of unity can be used to always extend a local Riemannian metric into a global one on whatever vector bundle above $X$. Thus, global Riemannian metrics always exist at our disposal. The downside is that they are all geodesically complete. Incapable to provide good classical models for cosmology and/or black hole physics. Unluckily, obstructions to extend a local Lorentzian metric into a global one are much more common. Paracompactness is enough for non-compact topologies. But on compact ones, it needs to be supplemented with the condition $\chi\left(X\right)=0$. Famously, even-spheres do not accept a Lorentzian structure.

The existence of metric structures on $X$ has interesting consequences for $FX$ and $P$. Via the Gram-Schmidt process, $g$ and $g'$ allow us to define in each $F_x X$ and $P_x$, respectively, subsets $F^\mathcal{O}_x X \subset F_x X$ and $P^\mathcal{O}_x \subset P_x$ of orthogonal frames. The disjoint union in all $x$ defines $F^\mathcal{O}X$ and $P^\mathcal{O}$. One can show that these do have the structure of embedded principal sub-bundles within $FX$ and $P$, respectively, with structure group $O\left(p,q \right) \; ; \; p+q=n$, if the metrics $g$ and $g'$ have signature $\left(p,q\right)$. If Riemannian ($p=0$), then the structure subgroup is $O\left(n\right)$. If Lorentzian ($p=1$), then $O\left(1,n-1\right)$. On orientable topologies, $GL\left(n,\mathbb{R}\right)$ can first be reduced to the orientation-preserving $GL^+\left(n,\mathbb{R}\right)$ (positive determinant), yielding $SO\left(p,q\right)$.

The proof of existence on paracompact $X$ rely on the quotient bundles $FX/O\left(p,q\right)$ and $P/O\left(p,q\right)$ admitting global sections, \textit{i.e.}, being trivial. This is true whenever the coset space $GL\left(n,\mathbb{R}\right)/O\left(p,q\right)$ is contractible. If $p=0$, this space is homotopic to $\mathbb{R}^n$. If $p=1$, it is homotopic to $\mathbb{R}P^{n-1}$ --- the $\left(n-1\right)$-dimensional real projective space. The former is clearly contractible, the latter is not. $\pi_1 \left(\mathbb{R}P^{n-1}\right) \simeq \mathbb{Z}_2$ and $\pi_k\left(\mathbb{R}P^{n-1}\right) \simeq \pi_k\left(S^{n-1}\right)$ for $k>1$. This is the bundle-theoretical reason why Riemannian structures always exist while Lorentzian ones do not.

Regardless of signature, whenever a metric structure exists, the embeddings $F^\mathcal{O}X \rightarrow FX$ and $P^\mathcal{O}\rightarrow P$ also do. Ultimately, this is the justification, omitted from Section~\ref{sec:holonomic_and_non-holonomic_frames}, that allowed us to extend the transformation group of non-holonomic frames from $SO\left(1,n-1\right)$ to $GL\left(n,\mathbb{R}\right)$: we assumed the existence of $\eta$. From a physical standpoint, the converse interpretation is promising. One can argue that whenever the quantum structure of spacetime changes to that of a smooth manifold $X$ with appropriated topology, then a corresponding symmetry breaking $GL\left(n,\mathbb{R}\right) \rightarrow O\left(p,q\right)$ occurs in $FX$ and $P$. This gives a comprehensive scenario in which metric structures arise dynamically, as Higgs-Goldstone type of fields~\cite{ivanenko1983a,nikolova1984,sardanashvily2016a}. Theories of induced gravity employing similar mechanism have been extensively explored in~\cite{macdowell1977a,stelle1979,gotzes1989,neeman1987,kirsch2005,leclerc2006,tresguerres2008,randono2010,sobreiro2011,mielke2011a,sobreiro2017a,sadovski2015}.

It should be apparent that to have a bundle of orthogonal frames by no means equates to have an everywhere flat Minkowski structure $\eta$. And, to realize the equivalence principle, we do need plug $FM$ onto $X$. $FM$ is first conceived over the Minkowski space $M$. The latter is paracompact and non-compact, thus a global Lorentzian metric $g'$ definitely exists on it. Further, $M$ is homotopic to $\mathbb{R}^n$ and thus contractible. This means that any bundle over it is trivial. Consequentially, $\omega'$ can be chosen as the canonical flat connection. By postulating that this connection is torsionless and metrical, we arrive at the Riemannian hypothesis that states that $\omega'$ is derived from the metric $g'=\eta$.

The existence of $\eta$ guarantees the existence of the global Lorentz frame $\tau_a$. This is realized via the embedding $F^\mathcal{O} M \rightarrow FM$, in which $SO(1,n-1)$ is structure subgroup. In $\tau_a$, $\eta$ assumes its well-known diagonal form $\eta\left(\tau_a,\tau_b\right)=\eta_{ab}\equiv \mathrm{diag}\left(-1, +1, \ldots, +1\right)$. By plugging $FM$ onto $X$ via the projection $\pi'$, we essentially localize on $X$ the global Minkowskian structures just mentioned. In summary, the equivalence principle is geometrically encoded in the following diagrams:
\begin{equation}
  \label{dia:p-bundle_ep}
  \begin{tikzcd}
    GL\left(n,\mathbb{R}\right)
    \arrow[r,hook]
    &
    FX
    \arrow[r,"h"]
    \arrow[d, "\pi" left]
    &
    FM
    \arrow[r,"q"]
    \arrow[d, "\pi'"]
    &
    F^\mathcal{O}M
    \arrow[d, "\pi''"]
    &
    \arrow[l,hook']
    SO(1,n-1)
    \\
    &
    X
    \arrow[r,"{\mathds{1}_X}"]
    &
    X
    \arrow[r,"{\mathds{1}_X}"]
    &
    X
    &
  \end{tikzcd} \tag{Diagram 5}
\end{equation}
or, equivalently, in terms of vector bundles,
\begin{equation}
  \label{dia:v-bundle_ep}
  \begin{tikzcd}
    GL\left(n,\mathbb{R}\right)
    \arrow[r,hook]
    &
    TX
    \arrow[r,"e"]
    \arrow[d, "\pi_{TX}" left]
    &
    TM
    \arrow[r,"q'"]
    \arrow[d, "\pi_{TM}"]
    &
    T^\mathcal{O}M
    \arrow[d, "\pi_{T^\mathcal{O}M}"]
    &
    \arrow[l,hook']
    SO(1,n-1)
    \\
    &
    X
    \arrow[r,"{\mathds{1}_X}"]
    &
    X
    \arrow[r,"{\mathds{1}_X}"]
    &
    X
    &
  \end{tikzcd} \tag{Diagram 6}\;,
\end{equation}
where $q$ and $q'$ are bundle contractions. Clearly, it is the existence of the bundle isomorphism $h$ --- or, equivalently, $e$ --- that allow us to formulate gravity on other bundles beyond natural ones. In Section~\ref{sec:on-shell_equivalence}, we proved that these different constructions are dynamically equivalent, in the classical realm, if the diagrams above hold true.

We are ready to state equations~\eqref{eq:field-transformation} in a geometrical fashion. They correspond to the pullback along $e$ of the non-holonomic metric $\eta$ and connection $A$ to the holonomic metric $g$ and connection $\Gamma$, respectively,
\begin{subequations}\label{eq:e-pullback}
  \begin{align}
    g      & = e^* \eta \;, \label{eq:e-pullback-of-nonholonomic-metric}  \\
    \Gamma & = e^* A \;, \label{eq:e-pullback-of-nonholonomic-connection}
  \end{align}
\end{subequations}
much in the spirit first presented in~\cite{giachetti1980}. Clearly, equations in~\eqref{eq:e-pullback} are a reincarnation of equation~\eqref{eq:soldered_connection}, but on an associated vector bundles, and including the metric field. Nevertheless, we again stress that, $e$ is not an isomorphism from $TX$ to itself, but an isomorphism from $TX$ to $TM$. Otherwise, $e$ would be the functorial lift of $lAut\left(X\right)$, according to~\ref{dia:categorical_lift}, and its matrix representation would be $\tensor{J}{^{\nu'}_\mu}(x)$. This is an important conceptual distinction that, in practice, only amount for a substitution from $\tensor{J}{^{\nu'}_\mu}(x)$ to $\tensor{e}{^A_\mu}(x)$ in the transformation law for tensor and connection fields. Since equations of motion should be covariant under the latter, it is also covariant under the former. This is what geometrically underpins our analytical result of Section~\ref{sec:on-shell_equivalence}.

Let $ U $ be a region such that
\begin{equation}\label{eq:degenerate}
  \det\left[\tensor{e}{^A_\mu}(x)\right]=0 \; \forall \; x \in U \;.
\end{equation}
Over such degenerate region, $ e $ is not an isomorphism. Thus, $TM$ cannot be realized as a bundle soldered to $TX$, as defined above. Physically, in $U$ the equivalence principle fails to hold.

%% file: content/sections/conclusion.tex
\section{Conclusions}\label{sec:conclusions}

In the present work, we extended the known classical equivalence between the non-holonomic $S_{\text{VHP}}\left[e,A\right]$ \textit{versus} the holonomic $S_{\text{EHP}}\left[g,\Gamma \right]$ theory of gravity, under field transformations~\eqref{eq:field-transformation}. The equivalence now holds in all spacetime dimensions, and in all metric-affine dynamics --- including arbitrarily high (but finite) high-derivative ones.

We presented a detailed geometric formulation of field transformations~\eqref{eq:field-transformation}, how they encapsulate the equivalence principle, and how their violation might break the equivalences aforementioned. Physically, this break equates to a scenario in which a non-holonomic gauge description of gravity is completely dissociated from spacetime; the internal degrees of freedom are not mimicking the external ones.

A known case in the literature is on degenerate spacetime regions. At them, the vector bundle morphism $e$ is, at most, surjective or injective. The \textit{vielbein} field is non-invertible or, in holonomic language, the metric tensor is singular. In~\cite{kaul2016a,kaul2016b,kaul2019}, it was shown that the classical equivalence between $S_{\text{VEP}}\left[e,A\right]$ and $S_{\text{EHP}}\left[g,\Gamma\right]$ breaks in these regions. Our analysis, in Section~\ref{sec:on-shell_equivalence}, allows to naturally extend this result to the generic metric-affine dynamics. Quantum mechanically, an earlier work by A.~A.~Tseytlin had already noticed this failure once $\det\left(\tensor{e}{^A_\mu}\right)=0$ configurations are allowed in the gravitational path integral~\cite{tseytlin1982}. In general, these spacetime regions are associated to topology-change~\cite{geroch1967,tipler1977,horowitz1991,borde1994,borde1999,heveling2022}.

Another way to violate~\eqref{eq:field-transformation} is to postulate the existence of a non-vanishing tensor field
\begin{equation}
  \label{eq:d-tensor}
  \tensor{D}{^A_{\mu\nu}} \equiv \partial_\mu \tensor{e}{^A_\nu} + \tensor{\Gamma}{^\alpha_{\mu\nu}}\tensor{e}{^A_\alpha}-\tensor{\omega}{^A_{B\mu}}\tensor{e}{^B_\nu}\;.
\end{equation}
In the literature,~\eqref{eq:d-tensor} is sometimes interpreted as the result of applying an exterior covariant derivative, defined on the spliced bundle $TX\times TM$, to $\tensor{e}{^A_\nu}$. Theories with non-vanishing $\tensor{D}{^A_{\mu\nu}}$ must live on the spliced bundle, and are concomitantly holonomic and non-holonomic. Thus, the question of equivalence becomes nonsensical. This scenario, however, presents a novel way to lift spacetime and gauge space symmetries into a single geometrical arena, finding recent applications in 11-dimensional supergravity, higher-spin gravity, and M-theory~\cite{hull2007,engquist2008,nicolai2014a,nicolai2014b}.

%% file: content/sections/acknowledgments.tex
\section*{Acknowledgments}

The author is thankful to R.~F.~Sobreiro and J.~Zanelli for the useful discussions during the development of this work.